# INTERPRETATION OF INTERACTION: A REVIEW

By Amy Berrington de González and D. R. Cox

*Johns Hopkins Bloomberg School of Public Health
and Nuffield College, Oxford*

Several different types of statistical interaction are defined and distinguished, primarily on the basis of the nature of the factors defining the interaction. Illustrative examples, mostly epidemiological, are given. The emphasis is primarily on interpretation rather than on methods for detecting interactions.

**1. Introduction.** Interaction is one of the fundamental concepts of statistical analysis. Establishing the presence or absence of interaction may be a key to correct interpretation of data. Discussion of interaction falls under three broad headings, namely, its definition, its detection and its interpretation. This paper is mostly devoted to the last, interpretation. Our illustrations are largely epidemiological; the relevance of the ideas is much wider.

We consider studies in which on a number of individuals there are observed one or more response (or outcome) variables and typically several explanatory variables, conveniently called factors, that are thought possibly to influence the response. We consider initially interaction between a given pair of factors. From the statistical perspective, interaction is said to occur if the separate effects of the factors do not combine additively. That is, interaction is a particular kind of nonadditivity. The terminology is in some ways unfortunate in that there is no necessary implication of, say, biological interaction in the sense of synergism or antagonism.

When the outcome is measured on a quantitative scale interaction on one scale may possibly be removed by a nonlinear transformation of the scale. For binary outcomes, representing say survival and death, interaction is defined via the nonadditivity of some function of the probability of death. When the probability is small, absence of interaction on the logistic scale implies that to a close approximation separate explanatory variables combine their effects multiplicatively. From a public health perspective, it may be preferable to consider instead or, as well, the probability scale itself when absence of interaction means additivity of effect [Berkson (1958)]. An interpretation via probabilities is then directly in terms of differences of numbers of individuals at risk.

Detection of interaction is achieved essentially by comparing the fits of models with and without interaction terms, or sometimes by estimation of defining parameters, and will hardly be discussed here; one of the main issues for choice, especially when one or both factors have several levels, concerns how general the interaction terms should be. That is, is it wise to restrict, initially at least, the interaction to particular patterns of effect? For a review of techniques for detecting interaction, see Cox (1984).

The paper begins by making an important distinction between types of explanatory variables. We then discuss a very simple situation not commonly thought of as illustrating interaction and then discuss the interpretation of the main types of two-factor interaction that can arise.

**2. Types of factor.** Factors, or explanatory variables, can be classified in various ways. First the levels of a factor may be defined by a quantitative variable, by an ordinal variable or the different levels may be qualitatively different. Examples are respectively dose level of medication, level of exposure (severe, moderate, absent) and centers (in a multi-center trial), when these are seen as essentially providing replication rather than as the focus of particular interest.

More importantly, for our purpose, we classify factors as:

- primary factors or what in some contexts might be called treatments or quasi-treatments,
- intrinsic factors defining the study individuals,
- nonspecific factors, representing groupings of the study individuals that are of no intrinsic interest but which may have nonnegligible effect on the response.

This classification is strongly context-specific.

In a randomized experiment the primary factors are those randomized treatments that form the focus of the study. In a comparable observational study they are broadly those that would have been treatments had randomization been feasible. Comparison of their effect aims at a causal interpretation, although in an observational study claims of causality have to be approached very cautiously. Conceptually, at least, for a given study individual, a primary factor might have been different from the value observed; thus, an individual might have been randomized to a different treatment from that actually encountered.

Intrinsic factors define the study individuals, and hence usually an individual could not have been randomized to receive a different "intrinsic factor." In an epidemiological context these typically include gender, socio-economic class, educational and family background. The role of many variables such as smoking status depends strongly on context; they may be a main focus of interest or be regarded as intrinsic. Genetic information about an individual may be taken as helping to define a study individual, and hence intrinsic, but in the study of a potentially Mendelian disease genetic information may be a primary factor. In the latter case we implicitly consider the question: what would the health status of this individual have been had this allele been different from how it is?

The two-factor interactions of most interest are those in which at least one factor is a primary factor and there are thus three main cases to consider. First, however, we discuss a simpler situation which at first sight may not seem to involve the concept of an interaction at all.

**3. Constancy of variance.** Consider a continuous response variable $y$ and, for simplicity, two treatments. In the absence of further structure in the data, we have a two-sample problem defined implicitly by two distribution functions $F_0(y)$ and $F_1(y)$ corresponding to the two treatments $T_0$ and $T_1$.

There is then a sense in which absence of interaction implies that one distribution is a translation of the other $F_1(y) = F_0(y - \theta)$.

This interpretation hinges on the notion of unit-treatment additivity. That is, the response observed on a particular individual is assumed to be the sum of a contribution characteristic of the individual and a constant defined by the treatment received. Whatever may be the distribution of the individual characteristics, this implies the stated translational form.

Thus, if $T_1$ is a potential cholesterol lowering drug and $T_0$ a control, absence of translational form would imply that on average the drug had a differential effect at different levels of cholesterol, on the scale in which cholesterol is measured.

There are now two cases. First, if two distribution functions $F_1(y)$ and $F_0(y)$ are such that as $y$ takes values over the support of the distributions $F_1(y) - F_0(y)$ takes both signs, then we say the distribution functions cross. If the distribution functions do not cross, it may be shown that a nonlinear transformation of $y$ induces translational form implying consistency with unit-treatment additivity on the new scale. If, on the other hand, the distribution functions do cross, clearly no such transformation is possible. In the illustrative example there would at least be the implication that $T_1$ is beneficial for some individuals and harmful for others.

If the distributions are approximately normal, they are characterized by means $(\mu_1, \mu_0)$ and variances $(\sigma_1^2, \sigma_0^2)$ and the distribution functions do not cross if and only if the variances are equal. Examination of equality of variance is quite commonly presented as a technical statistical issue concerned with the validity of tests of significance. It may often be more fruitful to consider it a substantive issue concerning implied interaction.

Now a normal distribution can at best be a good approximation and is unlikely to hold accurately in the extreme tails. Two normal distribution functions will cross at a probability level $\Phi(k)$, where

$$k = (\mu_1 - \mu_0)/(\sigma_1 - \sigma_0),$$

so that unless this is in a reasonably central part of the distribution, say, $|k| <$ 2, the crossing over is unlikely to have sensible substantive interpretation.

An approximate confidence band for the point of intersection can most readily be found by computing its profile likelihood function.

**4. Removable interaction.** We may call an interaction *removable* if a transformation of the outcome scale can be found that induces additivity. The importance of this is partly that presentation of the conclusions and the resulting interpretation may be improved by the resulting formal simplification. It would be a mistake, however, to achieve this simplification by measuring effects on a scale that is very hard to understand or interpret [Breslow and Day (1980)]. Note also that removable interactions are inconsistent with average effect reversal. For example, absence of interaction with gender on a transformed scale excludes the possibility that a treatment is on the average beneficial for men and on the average harmful for women, whatever the transformation of the measurement scale used.

For a continuous and positive response variable, $y$, the transformations commonly used are logarithmic and simple powers, occasionally with a translated origin. For binary data, the logistic or sometimes probit or complementary log scale may be effective. While achieving additivity of effects is helpful, interpretability is the overriding concern. Thus, the transformation from $y$ to $y^{1/3}$ might remove an interaction but, unless $y$ was a representation of a volume, $y^{1/3}$ might well not be a good basis for interpretation.

Terminology differs somewhat between fields of application; removable interactions are sometimes referred to as quantitative or ordinal interactions, where as nonremovable interactions are referred to as qualitative, cross-over or disordinal interactions [see Cronbach and Snow (1981)]. In the remainder of this paper we use the terminology of quantitative and qualitative interactions.

We now discuss and illustrate with examples the interpretation of the three main cases of interest, that is, interactions that involve a primary factor.

TABLE 1
*Estimated relative risks of lung cancer from Gustavsson et al. (2002)*

| Asbestos exposure | Current smoker | Relative risk | (95% CI) |
|---|---|---|---|
| No | No | 1.0 | – |
| No | Yes | 21.7 | (14.3, 32.6) |
| 2.5+ fiber-years | No | 10.2 | (2.5, 41.2) |
| 2.5+ fiber-years | Yes | 43.1 | (20.1, 88.6) |

## 5. Examples.

### 5.1. *Interaction between two primary factors.*

5.1.1. *Quantitative interaction.* Interpretation of quantitative interaction between two primary factors is complicated by the fact that, by definition, a quantitative interaction can be removed by transforming the scale of measurement. Results can be generalized more easily if the interaction is removed, but, as mentioned above, this should not usually be achieved at the expense of measuring effects on a scale that is difficult to interpret. Interpretation will often depend upon the aim of the investigation.

Gustavsson et al. (2002), for example, conducted a prospective study to investigate whether there was evidence of interaction between exposure to asbestos and smoking with respect to the risk of lung cancer. They performed two tests for interaction between these two primary factors: one for departure from an additive model and one for departure from a multiplicative model (equivalent to testing for additivity on the log scale). The relative risks for each exposure group compared to those subjects who were not exposed to either risk factor (noncurrent smokers who were not exposed to asbestos) are shown in Table 1.

The observed relative risk for the joint effect of the two risk factors (43.1) was significantly less than would have been expected under a multiplicative model ($21.7 \times 10.2 = 221.3$), but was slightly greater than expected under the additive model ($21.7 + 10.2 - 1 = 30.9$). However, departure from the additive model was not statistically significant. Hence, these results could either be interpreted as evidence that the effects of exposure to asbestos and tobacco could be additive with respect to the risk of lung cancer (i.e., act independently on this scale) or that there is a quantitative, sub-multiplicative interaction (i.e., they interact negatively) on a probability scale. Since biological or other information to support one scale over the other is rarely available [see Siemiatycki and Thomas (1981) for an example], it is not possible to choose between these two interpretations.

In this example the authors' aim was not to try to elucidate biological mechanisms but to inform policy. In particular, they were interested in whether special efforts should be made to help asbestos-exposed persons to stop smoking. Because the data were found to be consistent with the additive model for the joint effect of asbestos and smoking, this suggests that such a program is not necessary, as asbestos-exposed persons have approximately the same absolute increase in lung cancer risk from smoking as nonexposed persons. Several authors refer to this as absence of 'public health interaction' [Blot and Day (1979) and Rothman et al. (1980)].

TABLE 2
*Estimated relative risk of endometrial cancer in relation to cyclic-combined HRT use, according to body mass index [Beral et al. (2005)]*

| HRT use | Body mass index | Relative risk | (95% CI) |
| --- | --- | --- | --- |
| ever vs never | $<25$ kg/m$^2$ | 1.54 | (1.20, 1.99) |
| ever vs never | 25–29 kg/m$^2$ | 1.07 | (0.82, 1.40) |
| ever vs never | 30+ kg/m$^2$ | 0.67 | (0.49, 0.91) |

5.1.2. *Qualitative interaction.* Although it could be said that qualitative interaction is the only 'essential' statistical interaction, because it is nonremovable, if we use this approach, in practice, we would accept only effect reversal as evidence of interaction. Interesting and important quantitative interactions could therefore be over-looked. Qualitative interactions are relatively rare, but when they do occur they are usually of considerable interest. For example, in the Million Women UK cohort study there was evidence of qualitative interaction (effect reversal) between two primary factors: use of cyclic-combined hormone-replacement therapy (HRT) and body mass index, with respect to the risk of developing endometrial cancer [Beral et al. (2005)]. Women who were of normal body weight (body mass index $< 25$ kg/m$^2$) had a significantly increased risk of endometrial cancer if they had ever used this type of HRT, whereas women who were obese (body mass index of 30+ kg/m$^2$) had a significantly reduced risk of endometrial cancer if they had ever used this type of HRT compared to never users. A formal test should usually be performed to assess whether the qualitative interaction could be due to chance variation; see, for example, Azzalini and Cox (1984).

Note that the approach used to analyze and display the data will impact on the interpretation. The approach of a single baseline group (Table 1) allows for easy examination of the consistency with different models, such as the additive versus the multiplicative model, but does not reveal immediately whether there is qualitative interaction. The opposite is true for the approach of multiple contingency tables (Table 2).

5.1.3. *Continuous scale interactions.* Some special considerations apply in considering interaction between two primary factors both with levels spec-ified quantitatively. An example would concern the levels of two different atmospheric pollutants, the outcome being some measure of disease incidence. For given levels of other explanatory variables, interaction between the two quantitative factors with levels $x_1$ and $x_2$ amounts to departure from the so-called generalized additive model [Hastie and Tibshirani (1990)]

$$E\{Y(x_1, x_2)\} = a_1(x_1) + a_2(x_2),$$

where $Y(x_1, x_2)$ is the outcome for an individual with the specified levels of the explanatory variables.

There are two broad situations. In one $x_1$ and $x_2$ are very different kinds of factors which may individually have effects on response that are quite complicated, but which may act virtually independently inducing additivity. In such a situation interaction would be tested formally by introducing a term, or possibly a small number of additional terms, into the model. These might, for example, be a simple product such as $x_1 x_2$ or possibly better, $\hat{a}_1(x_1)\hat{a}_2(x_2)$, where $\hat{a}_j(x_j)$ is a preliminary estimate of $a_j(x_j)$.

In a contrasting situation $(x_1, x_2)$ are coordinates specifying points in a factor space and other coordinate systems may possibly be more interpretable. A notion stemming from the industrial response surface literature is that in the absence of quantitative background knowledge it may be best to think of the expected response as a function of $(x_1, x_2)$ that within a restricted region can be expanded in a Taylor series around some central reference level. From this perspective, if a model linear in the explanatory variables is inadequate, it will be sensible to add terms in $(x_1^2, x_1 x_2, x_2^2)$, of which the middle one represents interaction. In this context the generalized additive model may not be reasonable; generality of the functions $a_j(x_j)$ combined with exclusion of product (interaction) terms would probably be justified only as a device for transforming the individual $x_j$ to some relatively simple form for which interpretation via a first-order model is available. For studies of behavior near a local stationary value, use of at least second-order terms is needed, absence of interaction would mean that the local quadratic approximation had principal axes along the coordinate axes and, in general, there seems to be no reason to expect this. In such situations it may be best to abandon the main effect-interaction framework as a basis for interpretation and to concentrate on the expected response as a function to be estimated in some hopefully enlightening form [Box and Draper (2007)].

5.2. *Interaction between a primary and intrinsic factor.* The interpretation of interaction between a primary and an intrinsic factor may be quite straightforward. A pattern of effects has to be studied to some extent separately at the different levels of the intrinsic factor; this is sometimes also referred to as examination of effect-modification. Typically, if interaction is present, the main effect of the primary factor, while it may sometimes provide a useful qualitative synthesis, is not relevant for detailed interpretation. It involves an averaging over levels of the intrinsic factor which may be essentially meaningless. However, if it is found that the main effect of the primary factor is stable across the levels of the intrinsic factor, this implies that the findings are more generalizable.

TABLE 3
*Estimated relative risk of Parkinsons disease in relation to coffee consumption, according to sex [Ascherio et al. (2004)]*

| Coffee consumption | Sex | Relative risk | (95% CI) |
|---|---|---|---|
| 6+ vs 0 cups/week | males | 0.34 | (0.16, 0.75) |
| 6+ vs 0 cups/week | females | 1.09 | (0.61, 1.93) |

Although the statistical methods for evaluating interaction between a primary and intrinsic factor are essentially the same as those for the evaluation of interaction between two primary factors, the route to interpretation is different, because the roles of the primary and the intrinsic factor are asymmetrical.

If the intrinsic factor has quantitative levels, more elaborate models may aid interpretation. In these the nature of an interaction may change smoothly, or indeed linearly, with the level of the intrinsic variable.

For example, Ascherio et al. (2004) found evidence that high coffee consumption was associated with a significantly reduced risk of Parkinson's disease for men, but there was no evidence of such an effect for women (Table 3). Hence, it is not appropriate to summarize these results without reference to sex. The average risk of Parkinson's disease from high level coffee consumption for men and women combined would be meaningless. The asymmetry between the primary and the intrinsic factor can be understood here by considering what the interpretation would be if they had presented the relative risk of Parkinson's disease associated with sex according to level of coffee consumption. This is clearly not a sensible biological viewpoint.

5.3. *Interaction between a primary and a nonspecific factor.*  Suppose for simplicity of discussion that there are two alternative treatments $T$ and $C$ and that an estimate of the treatment contrast can be found separately at a number of centers, these being regarded as defining nonspecific factors in the sense explained above.

Two rather different situations need consideration. In one an internal estimate of the precision of these individual contrasts is available, either from explicit replication within centers or from implicit replication, for instance, a reasoned assumption of binomial or Poisson variability. If the treatment by center interaction is appreciable and clearly statistically significant, there

INTERPRETATION OF INTERACTION: A REVIEW  9is unexplained additional variation present affecting the primary treatment contrast. This should be explained if at all possible, for example, by regression on whole-center features.

If that is not possible, it may be unavoidable to treat the additional variation as random and to introduce an additional component of variance. The presence of this component will inflate the standard error of the primary treatment contrast, and, unless the centers contribute essentially equal amounts of information, will move the weighting to be attached to the different centers in the direction of equal weighting. The implicit treatment contrast of concern is now an average over an ensemble of repetitions. Note that if the degrees of freedom available to estimate this additional component of variance are small, estimation of it, while formally possible, is extremely fragile and it is likely to be wiser either simply to list estimates center by center or to use a sensitivity analysis of dependence on the poorly estimated component.

The inclusion of an additional component of variance will typically inflate, possibly appreciably, the estimated standard error of the overall effect. Such an analysis is often described as treating centers as a random effect. This is a little misleading, however, in that centers are unlikely to be a random sample from a meaningful population. Rather, it is the unexplained interaction that is being modeled as generated stochastically.

If, however, there is no effective replication within centers then the treatment by center interaction provides a base for error estimation; the simplest special case is the standard analysis of a randomized block design.

Duijts et al. (2003), for example, conducted a meta-analysis of epidemiological studies of stressful life events and the risk of breast cancer. When the results from all eleven published epidemiological studies were combined the summary odds ratio for ever versus never having had a stressful life event was 1.77 (95% CI: 1.31 to 2.40). However, there was evidence of significant heterogeneity between the results from the eleven studies (i.e., interaction with the nonspecific factor 'study'). The authors investigated whether several study level primary and intrinsic factors might explain this between study heterogeneity. The results in Table 4 show that the summary odds ratios were found to vary significantly according to whether there had been adjustment for the key confounding factors ($p < 0.001$). Between the studies that had adjusted for the key confounding factors there was still, however, significant heterogeneity that could not be explained by other study level factors. This additional heterogeneity, having no known deterministic explanation, was then treated as random and incorporated as an additional component of variance using a random effects model.

The use of the random effects model implicitly allows for the possibility of qualitative interactions between the primary and nonspecific factor. Some have argued against the use of this approach, because, as noted earlier, qualitative interactions should be relatively uncommon [Peto (1982)]. There are in any case substantial difficulties in combining studies where the supplementary variables used to adjust, say, the odds ratio, are very different for the distinct studies. More generally, the conceptual difficulties in treating replication in space or time as random were clearly set out in one of the earliest treatments of the summarization of evidence from repeated studies [Yates and Cochran (1938)].

TABLE 4
*Estimated summary odds ratios for breast cancer and stressful life events, according to confounding adjustment [Duijts et al. (2003)]*

| Stressful life events | Adjusted for key confounders? | Odds ratio | (95% CI) |
|---|---|---|---|
| yes vs no | no | 1.04 | (0.90, 1.20) |
| yes vs no | yes | 2.22 | (1.39, 3.56) |

5.4. *Higher-order interaction.* The difficulty of interpreting interactions increases rapidly with the number of factors involved, even if, in principle, the points made in connection with two-factor interactions cover many of the ideas needed. For example, Znaor et al. (2003) conducted a study of risk factors for oral cancer in Indian men. There was evidence that the joint effect of the three primary risk factors of interest (tobacco smoking, tobacco chewing and alcohol drinking) was approximately additive, but was significantly less than multiplicative (additive on the log-scale). Interpretation of the source of the sub-multiplicative three-way interaction can be aided by investigation of its source. Table 5 shows the odds ratios for each combination of the three risk factors compared to those that were not exposed to any of the three factors. The observed odds ratio for the joint effect of all three risk factors (16.34) was significantly less than would have been expected under the multiplicative model (58.14). Examination of each of the two-factor interactions shows that the joint effect of smoking and chewing tobacco was much lower than would have been expected under the multiplicative model (8.53 compared to 22.71). The joint effect of smoking and alcohol was also slightly lower than expected (4.81 compared to 6.27), but the observed joint effect of chewing tobacco and alcohol was consistent with the expected multiplicative joint effect (24.28 compared to 23.73). Hence, the main source of the sub-multiplicative three-way interaction appears to be the sub-multiplicative two-way interaction between smoking and chewing tobacco, but the sub-multiplicative interaction between smoking and alcohol may have contributed also. For binary data a formal test of 3 factor interactions in a $2 \times 2 \times 2$ table was given by Bartlett (1935).

TABLE 5
*Odds ratio (OR) for interaction for combinations of smoking, chewing tobacco and alcohol for the risk of oral cancer [Znaor et al. (2003)]*

| Smoking | Chewing tobacco | Alcohol | Odds Ratio (and 95% CI) |
|---|---|---|---|
| No  | No  | No  | 1.00 (–) |
| No  | Yes | No  | 9.27 (6.79–12.66) |
| Yes | No  | No  | 2.45 (1.94–3.10) |
| No  | No  | Yes | 2.56 (1.42–4.64) |
| Yes | Yes | No  | 8.53 (6.13–11.89) |
| No  | Yes | Yes | 24.28 (14.87–39.65) |
| Yes | No  | Yes | 4.81 (3.74–6.19) |
| Yes | Yes | Yes | 16.34 (12.13–22.00) |

In the previous discussion we have not suggested interpretations directly based on the formal parameters used in representing interactions in a model, regarding such models as more useful for testing for interaction than for its interpretation. In some applications, however, the pattern of, say, two-factor interactions, may be of prime concern. The stability of that pattern, for example, over replication of a nonspecific factor is then of interest.

An example is the study of social mobility where the primary data are essentially square contingency tables with the rows labeled by class of origin and the columns by class of destination. Interest may lie not in the changes in the marginal distribution between origin and destination, but rather in the pattern of interactions and in the stability of that pattern across time or countries.

This can be represented as follows. In one study let $\pi_{ij}$ be the probability that an individual is in origin class $i$ and destination class $j$. Write

$$\pi_{ij} = \pi_{i.}\pi_{.j}\psi_{ij},$$

where $\pi_{i.}, \pi_{.j}$ are marginal probabilities and the $\psi_{ij}$ satisfy the appropriate constraints. Now suppose that there is a third factor, say, a nonspecific factor. When this takes level $k$, we write the corresponding probability $\pi_{ij;k}$; that is, for each fixed $k$, this defines a probability distribution over the corresponding square table. Then a model in which the pattern of interaction is essentially the same for each level of $k$ but the magnitude of the interaction effect varies is represented in the form

$$\pi_{ij;k} = \pi_{i.;k}\pi_{.j;k}\rho_k\psi_{ij}.$$

This is one of a quite wide range of special models that can be considered for multiple contingency tables [Agresti (1990) and Goodman (1985)]. We do not discuss here the directly related, although conceptually different, literature of interaction in multiple contingency tables in which the different dimensions of the table are treated on an equal footing. The connection between log linear models and additive models [Lancaster (1969) and Darroch 1974] parallels the present discussion.

A rather different aspect of higher-order interaction for binary observations concerns the possible reversal of association as between marginal and conditional association, the Yule–Simpson effect [Yule (1903)]. A related issue is the possibility of spurious allelic association [Cardon and Palmer (2003)] where an observed dependence arises from mixing individuals, say, from different ethnic groups within each of which independence holds. This in turn is related to latent class analysis [Lazarsfeld and Henry (1968)] in which the aim is to represent observed multivariate dependencies by a small set of latent classes within each of which independence holds. A quantitative discussion of the modifying effect of marginalizing in this context is given by Cox (2003).

**6. Epistasis.** We return to the relatively simple situation in which we concentrate on a two-way table showing the mean response at various levels of two factors, at least one a primary factor. Our primary route to interpretation is via the notion of the no-interaction model as a reference model with departures from it, if they are present, described essentially verbally. There are, however, other possible representations that are in a sense just as simple as the no-interaction model. In genetics these are described as epistasis; different authors use the term somewhat differently.

Suppose, for simplicity, that there are two two-level factors specified by $i = -1, 1$ and $j = -1, 1$ and that the mean response at level $(i, j)$ is

$$\mu_{ij} = \mu + \alpha i + \beta j + \gamma ij.$$

Then the no-interaction model has $\gamma = 0$.

One simple epistatic model has

$$\mu_{11} = \nu + \lambda, \qquad \mu_{ij} = \nu \qquad \text{(otherwise)}.$$

This is a two-parameter model, as contrasted with the three parameter no-interaction model. Yet the epistatic model is not a special case of the no-interaction model. The totally null case $\lambda = 0$ is typically of no interest in this context and we assume that the data strongly exclude this.

Comparison of the models is most fruitfully achieved by testing separately consistency with the two models, leading to the conclusion, assessed by $p$-values, that the data are consistent with one, both or neither model.

We deal in outline with the simplest case of normally distributed data with equal sample sizes and constant variance but the details are not essentially different if, for instance, the data are represented by logistic models for probabilities.

Consistency with the no-interaction model can be tested only in effect by the least-squares estimate of $\gamma$ in the full model. Consistency with the epistatic model is tested by the mutual consistency of the three means excluding $\mu_{11}$ leading to a variance-ratio test with upper degrees of freedom equal to two. Unless there is further information, such as that the two factors are expected to have approximately equal effects of the same sign, $\alpha = \beta$, there is no basis for extracting a single degree of freedom.

Parallel tests based on the relevant log likelihood functions are available more generally.

**7. Interaction in balanced factorial designs.** Historically many of the ideas about interaction were first formulated in detail in connection with randomized factorial experiments, including those of quite complicated form. For such factorial experiments, at least those with a continuous and approximately normally distributed outcome, the powerful technique of analysis of variance allows the simultaneous inspection of interactions of all orders. Moreover, the distinction between factors describing the structure of the experimental units, block factors, and those determining the randomized treatments corresponds to the distinction between intrinsic and nonspecific factors contrasted with primary factors.

The role of analysis of variance in such contexts is partly in establishing via the table of degrees of freedom the logical structure of the data, and partly in indicating how the error to be attached to any type of contrast is to be estimated. This last is particularly important when treatments and experimental units have relatively complicated structure and lead to different sources of error, all based in effect on interactions between treatment and components of nonspecific variation.

In the absence of special reasons to the contrary, it will be sensible to start the formal analysis of such data by finding the full analysis of variance table together with all two- and some three-way tables of means and associated standard errors. This involves typically calculation of interactions of many different orders. Significance of many interactions involving, in particular, an intrinsic factor often suggests splitting the data into separate sections on the basis of that factor, for example, analyzing male and female sections separately. Use of other than the full analysis of variance table, or in other words, pooling of terms, may be needed to enhance error estimation, but this is to be regarded as a second-order effect.

The special feature of analysis by the standard normal-theory linear model is that the decomposition of the observational vector into orthogonal components, and therefore the additivity of sums of squares, typically allows assessment of effects of all orders virtually simultaneously. Analogous procedures, for example, log likelihood decompositions, are available for more general models and unbalanced data, but are typically contingent on a full model specification. That is, omission of certain terms from a model changes estimates of the other parameters. This tends to make an approach starting from a very general model with many interaction terms impracticable in such situations. It is the analysis strategy for detecting interactions that is changed rather than any issue of interpretation.

**8. Interaction detection in relatively large systems.** The emphasis in this paper is on the interpretation of interactions rather than on their detection, but we now comment briefly on interaction-detection in analyses in which the primary emphasis is on the representation of dependency of outcome on a fairly large number of explanatory variables. This is often in the first place specified by some form of linear regression representing main effects of the explanatory variables, in particular, identifying those with major effects on the outcome. It will be essential in interpreting such relations to distinguish between the various kinds of explanatory factors and to ensure that the relation fitted is consistent with any internal structure among the primary explanatory variables [Cox and Wermuth (1996)].

Subject to that, a search for interactions among the explanatory variables, will often be confined to interaction involving at least one primary factor. In some cases it may be feasible to fit all such interactions simultaneously, as, for example, in the previous section. More commonly, in large observational studies it is likely to be preferable to fit relevant interactions as single degrees of freedom at a time and to make a normal probability plot from the resulting $t$ statistics [Cox and Wermuth (1994)].

**9. Ill-specified interactions.** It has been implicit in the previous discussion that each interaction of potential interest can be encapsulated if not in a single parameter at least in a very small number. This is desirable for, among other reasons, incisive interpretation. This fails if, for example, the data are essentially, after adjustment for other effects, in the form of an $r \times c$ table suggesting an interaction test having $(r-1)(c-1)$ degrees of freedom. If one or both $r$ and $c$ are not small, the resulting procedure has some sensitivity against a general class of departures from additivity, but poor properties for specific kinds of departure which may have special plausibility. One route is to take an interaction defined by the product of scores attached separately to the rows and columns. In the absence of scores derived, for example, from the ordinal character of the levels, products of estimated main row and column effects may be used [Tukey (1949)]. See also Yates (1948).

It is a matter of context whether importance lies primarily in establishing and interpreting interaction or in showing its effective absence. Absence of interpretable interaction of an important primary factor with intrinsic and nonspecific factors is a partial base for hope that any conclusion is

INTERPRETATION OF INTERACTION: A REVIEW 15

generalizable to new situations and applicable to specific individuals. One of the broad themes of the paper is that the importance of the notion of interaction is in no way confined to relatively complicated issues connected with multiple contingency tables and complex factorial experiments.

## REFERENCES

Agresti, A. (1990). *Categorical Data Analysis*. Wiley, New York. MR1044993
Ascherio, A., Weisskopf, M. G., O'Reilly, E. J., McCullough, M. L., Calle, E. E., Rodriguez, C. and Thun, M. J. (2004). Coffee consumption, gender, and Parkinson's disease mortality in the Cancer Prevention Study II cohort: The mortality effects of estrogen. *Am. J. Epidemiology* **160** 977–984.
Azzalini, A. and Cox, D. R. (1984). Two new tests associated with analysis of variance. *J. Roy. Statist. Soc. Ser. B* **46** 335–343. MR0781894
Bartlett, M. S. (1935). Contingency table interactions. *J. Roy. Statist. Soc. Ser. B* **2** 248–252.
Beral, V., Bull, D., Reeves G. and Million Women Study Collaborators (2005). Endometrial cancer and hormone-replacement therapy in the Million Women Study. *Lancet* **365** 1543–1551.
Berkson, J. (1958). Smoking and lung cancer: Observations on two recent reports. *J. Amer. Statist. Assoc.* **53** 28–38.
Blot, W. J. and Day, N. E. (1979). Synergism and interaction: Are they equivalent? *Am. J. Epidemiology* **110** 99–100.
Box, G. E. P. and Draper, N. R. (2007). *Response Surfaces, Mixtures and Ridge Analyses*, 2nd ed. Wiley, New York. MR2293880
Breslow, N. E. and Day, N. E. (1980). *Statistical Methods in Cancer Research.* **1**. *The Analysis of Case-Control Studies.* IARC Scientific Publications, Lyon, France.
Cardon, L. R. and Palmer, L. J. (2003). Population stratification and spurious allelic association. *Lancet* **361** 598–604.
Cox, D. R. (1984). Interaction. *Internat. Statist. Rev.* **52** 1–31. MR0967201
Cox, D. R. (2003). Conditional and marginal association for binary random variables. *Biometrika* **96** 982–984. MR2024772
Cox, D. R. and Wermuth, N. (1994). Tests of linearity, multivariate normality and the adequacy of linear scores. *Appl. Statist.* **43** 347–355.
Cox, D. R. and Wermuth, N. (1996). *Multivariate Dependencies.* Chapman and Hall, London. MR1456990
Cronbach, L. J. and Snow, R. E. (1981). *Aptitudes and Instructional Methods*: *A Handbook for Research on Interactions.* Irvington Publishers Inc, New York.
Darroch, J. N. (1974). Multiplicative and additive interaction in contingency tables. *Biometrika* **61** 207–214. MR0403087
Duijts, S. F. A., Zeegers, M. P. A. and Borne, B. Vd. (2003). The association between stressful life events and breast cancer risk: A meta-analysis. *Internat. J. Cancer* **107** 1023–1029.
Goodman, L. A. (1985). The analysis of cross-classified data having ordered and/or unordered categories: Association models, correlation models, and asymmetry models for contingency tables with or without missing entries. *Ann. Statist.* **13** 10–69. MR0773152
Gustavsson, P., Nyberg, F., Pershagen, G., Scheele, P., Jakobsson, R. and Plato, N. (2002). Low dose exposure to asbestos and lung cancer: Dose-response relations and interaction with smoking in a population-based case-referent study in Stockholm, Sweden. *Am. J. Epidemiology* **155** 1016–1022.

Hastie, T. J. and Tibshirani, R. J. (1990). *Generalized Additive Models.* Chapman and Hall, London. [MR1082147](MR1082147)

Lancaster, H. O. (1969). *The Chi-Squared Distribution.* Wiley, New York. [MR0253452](MR0253452)

Lazarsfeld, P. F. and Henry, N. W. (1968). *Latent Structure Analysis*. Houghton-Mifflin, Boston.

Peto, R. (1982). Statistical aspects of cancer trials. In *Treatment of Cancer* (K. E. Halnan, ed.) 867–871. Chapman and Hall, London.

Rothman, K. J., Greenland, S. and Walker, A. M. (1980). Concepts of interaction. *Am. J. Epidemiology* **112** 467–470.

Siemiatycki, J. and Thomas, D. C. (1981). Biological models and statistical interactions: An example from multistage carcinogenesis. *Internat. J. Epidemiology* **10** 383–387.

Tukey, J. W. (1949). One degree of freedom for non-additivity. *Biometrics* **5** 233–242.

Yates, F. (1948). The analysis of contingency tables with groupings based on quantitative characters. *Biometrika* **35** 176–181.

Yates, F. and Cochran, W. G. (1938). The analysis of groups of experiments. *J. Agric. Sci.* **28** 556–580.

Yule, G. U. (1903). Notes on the theory of attributes in statistics. *Biometrika* **2** 121–134.

Znaor, A. A., Brennan, P., Gajalakshmi, V., Mathew, A., Shanta, V., Varghese, C. and Boffeta, P. (2003). Independent and combined effects of tobacco smoking, chewing and alcohol drinking on the risk of oral, pharyngeal and esophageal cancer in Indian men. *Internat. J. Cancer* **105** 681–686.

Johns Hopkins Bloomberg
 School of Public Health
615 N Wolfe St
Baltimore, Maryland 21205
USA
E-mail: [aberring@jhsph.edu](aberring@jhsph.edu)

Nuffield College
New Road
Oxford, OX1 1NF
United Kingdom
E-mail: [david.cox@nuffield.ox.ac.uk](david.cox@nuffield.ox.ac.uk)